\pdfoutput=1
\documentclass[11pt]{article}

\usepackage{EMNLP2022}

\usepackage{times}
\usepackage{latexsym}

\usepackage[T1]{fontenc}
\usepackage[utf8]{inputenc}

\usepackage{microtype}

\usepackage{inconsolata}
\usepackage{graphicx,booktabs,multirow,amsmath,bm,amssymb}
\usepackage{comment}

\title{Speaker Overlap-aware Neural Diarization for Multi-party \\Meeting Analysis}

\author{Zhihao Du \and Shiliang Zhang \and Siqi Zheng \and Zhijie Yan \\
	    Speech Lab, Alibaba Group, China \\
        \texttt{\{neo.dzh,sly.zsl\}@alibaba-inc.com}}

\begin{document}
\maketitle
\begin{abstract}
Recently, hybrid systems of clustering and neural diarization models have been successfully applied in multi-party meeting analysis.
However, current models always treat overlapped speaker diarization as a multi-label classification problem, where speaker dependency and overlaps are not well considered.
% Our system comprises a spectral clustering module to extract speaker profiles and a neural diarization model to deal with overlapping speech.
To overcome the disadvantages, we reformulate overlapped speaker diarization task as a single-label prediction problem via the proposed power set encoding (PSE).
Through this formulation, speaker dependency and overlaps can be explicitly modeled.
To fully leverage this formulation, we further propose the speaker overlap-aware neural diarization (SOND) model, which consists of a context-independent (CI) scorer to model global speaker discriminability, a context-dependent scorer (CD) to model local discriminability, and a speaker combining network (SCN) to combine and reassign speaker activities.
Experimental results show that using the proposed formulation can outperform the state-of-the-art methods based on target speaker voice activity detection, and the performance can be further improved with SOND, resulting in a 6.30\% relative diarization error reduction.
% explicitly modeling speaker dependency and overlaps brings a 10.8\% relative improvement. While CD scorer affects diarization performance, CI scorer improves the robustness to noises in speaker profiles. 
% Besides, we find SCN is more effective than other commonly-used network architectures.
% As a result, the proposed system outperforms the state-of-the-art methods with .
\end{abstract}

\section{Introduction}
Speaker identity is an import information for speaker-attributed automatic speech recognition \cite{CarlettaABFGHKKKKLLLMPRW05,BarkerWVT18} and dialogue comprehension \cite{HeTL021}, especially in multi-party meeting scenarios \cite{YuZFXZDHGYMXB22}.
Recently, speaker diarization is developed to detect and track speakers with acoustic features.
In general, speaker diarization methods can be divided into three groups, i.e., clustering-based algorithms, end-to-end models and hybrid systems.

Clustering-based methods comprise three individual parts, including speech segmentation, embedding extraction, and clustering algorithm.
Specifically, a long-term audio is first split into several segments by voice activity detection (VAD). 
Then, speaker embeddings \citep{DehakKDDO11, WanWPL18, SnyderGSPK18} are extracted for each segment, which are partitioned into clusters by unsupervised clustering algorithms, such as k-means \citep{DimitriadisF17}, spectral clustering \citep{NingLTH06}, agglomerative hierarchical clustering (AHC) \citep{Garcia-RomeroSS17}, and Leiden community detection \citep{ZhengSQ2022}.
However, such clustering algorithms are in an unsupervised manner, which does not minimize the diarization errors directly.
To overcome this issue, neural models are introduced, such as \citet{LiK0W21} and \citet{ZhangSWYY22}.
Recently, the variational Bayesian hidden Markov model (VBx) is introduced to refine the clustering results \citep{DiezBWRC19}.
Although VBx achieves impressive performance on several datasets \citep{CastaldoCDLV08, RyantCCCDGL19}, it has trouble handling overlapping speech due to speaker-homogeneous assumption.

End-to-end (E2E) approaches treat speaker diarization as a multi-label classification problem, where a deep neural network is employed to predict a set of binary labels (speaker activities) for each timestep. 
In this way, E2E models can deal with overlapping speech and minimize diarization errors directly.
In \citet{FujitaKHNW19}, the utterance-level permutation-invariant training (uPIT) loss \citep{KolbaekYTJ17} is employed to train the end-to-end neural diarization (EEND) model which is further improved in \citet{FujitaKHXNW19} and \citet{Liu2021Conformer}.
To deal with an unknown number of speakers, the encoder-decoder based attractor is involved into EEND \citep{HoriguchiF0XN20}.
Another E2E model is the recurrent selective attention network (RSAN), which jointly performs source separation, speaker counting, and diarization.
Although a lot of efforts have been made \citep{HoriguchiWGXTK21,XueHFTWPN21,WangL22b}, E2E models still have a low scalability to a large number of speakers and long-term audios due to the label permutation problem and memory limitation. 
% ZhangSWYY22,LiK0W21
To deal with long-term audios in multi-party meeting scenarios, a hybrid system of clustering and neural diarization model can be a reasonable approach \citep{MedennikovKPKKS20,KinoshitaDT21,Yu2022Summary}.
In \citet{MedennikovKPKKS20}, the unconstrained k-means algorithm is used to extract x-vectors \citep{SnyderGSPK18} as speaker profiles. 
Then, the profiles are consumed by a neural diarization model, target-speaker voice activity detection (TSVAD), to obtain frame-level diarization results.
TSVAD model takes acoustic features along with x-vectors for each speaker as inputs. A set of binary classification output layers produces activities of each speaker.
Recently, TSVAD is widely-used in multi-party meeting scenarios, and achieves the state-of-the-art performance \citep{Yu2022Summary,Weiqing2022,ZhengLWMKWWSM22}.

In both TSVAD and E2E approaches, overlapping speaker diarization is formulated as a multi-label classification problem.
This formulation ignores speaker dependency, and speaker overlaps are not explicitly modeled, which can cause much miss detection and false alarm errors.
To overcome the disadvantages, we reformulate overlapping speaker diarization as a single-label prediction problem via power set encoding. 
In this way, speaker dependency and overlaps are explicitly modeled by predicting a single label of speaker combinations rather than a set of multiply binary labels.
To fully leverage this formulation, we further propose the speaker overlap-aware neural diarization (SOND) model. 
In SOND, a context-independent (CI) scorer is employed to model global speaker discriminability, while a context-dependent (CD) scorer is involved to discover local discriminability of speakers in the same audio. 
Given CI and CD scores, a speaker combining network is proposed to combine and reassign speaker activities.
As a results, the proposed method achieves a 6.30\% relative improvement than the state-of-the-art methods on a challenging real meeting scenarios \footnote{Our code is available at https://github.com/ZhihaoDU/ du2022sond}.

% Our contributions are three-fold. First, we reformulate speaker diarization from a multi-label classification problem to a single-label prediction problem, which has not been studies in previous literature.
% Second, we propose a novel neural diarization model, SOND, which outperforms the state-of-the-art model, TSVAD.
% Third, we refine the hybrid system  by improving 

% \section{Related Work}

\begin{figure*}[t!]
	\centering
	\includegraphics[width=0.81\linewidth]{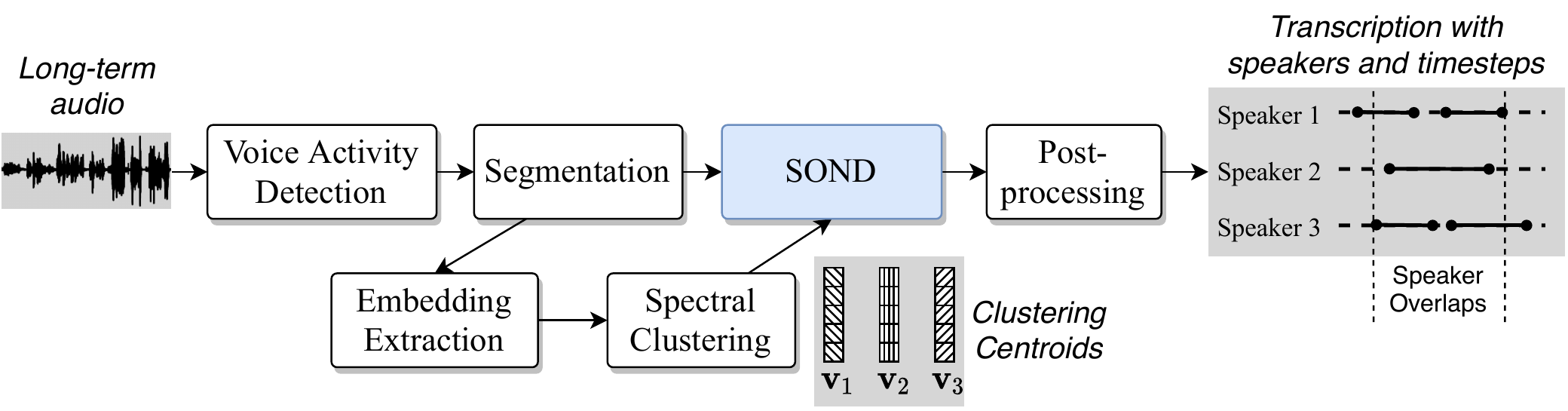}
	\caption{An overview of our speaker diarization system. SOND denotes the proposed speaker overlap-aware neural diarization model.}
	\label{fig:overview}
\end{figure*}
\section{System Description}
An overview of our speaker diarization system is shown in Figure \ref{fig:overview}. 
Given a long-term audio, we first remove its silence regions according to the results of voice activity detection. 
Then, we uniformly clip the voiced signal into segments with a fixed length. 
For each segment, a speaker embedding is extracted from a pre-trained neural network.
Next, we perform spectral clustering on the extracted speaker embeddings, and the clustering centroids are treated as speaker profiles.
Given the clipped segments and speaker profiles as inputs, a neural diarization model is employed to estimate speaker activities across all timesteps.
Finally, we perform post-processing on the diarization results of segments and obtain the transcription of entire long-term audio.

\subsection{Voice Activity Detection}
Voice activity detection (VAD) aims at finding out the voiced regions in an audio signal.
In multi-party meeting scenarios, regions with one or more speakers are marked as ``voiced'', and others are treated as ``unvoiced''.
In our experiments, VAD results have been already provided by the corpus for fair evaluation, therefore, we directly use the official VAD results in our system.
% In multi-party meeting scenarios, this task can be more challenging, since there can be more than one speaker talking at the same time. 

\subsection{Segmentation}
For neural diarization model, we uniformly clip the voiced signal with a window size of 16s shifted every 4s.
For embedding extraction, to guarantee the speaker-purity of each segment, we further clip the voiced segments into shorter chunks with a chunk size of 1.28s and a shift of 0.64s.

\subsection{Embedding Extraction}
\label{sec:spk_emb_ext}
We employ the ResNet34 \citep{HeZRS16} as our speaker embedding extractor. The encoding layer is based on statistic pooling (SP), and the dimension of the speaker embedding layer is 256. More details about model architecture are provided in \ref{sec:app:resnet34}. The ArcFace loss function \citep{DengGXZ19} with a margin of 0.25 and softmax pre-scaling of 8 is used to optimize the speaker embedding model.

\subsection{Spectral clustering}
In our system, we employ a bidirectional long and short term memory (Bi-LSTM) based recurrent neural network (RNN) to construct the affinity matrix for spectral clustering as in \citet{LinYLBB19}.
Specifically, $m$ extracted speaker embeddings $[\mathbf{e}_1,\mathbf{e}_2,\dots,\mathbf{e}_m]$ are concatenated with repeated $\mathbf{e}_i$ to predict the $i$-th row of affinity matrix.
 % $\mathbf{A}$ can be obtained by:
\begin{comment}
\begin{equation}
\begin{split}
\mathbf{A}_i &= [\mathbf{A}_{i,1},\mathbf{A}_{i,2},\dots,\mathbf{A}_{i,m}]\\
&=f(\left[\begin{array}{cccc}
\mathbf{e}_1 & \mathbf{e}_2 & \dots & \mathbf{e}_m \\
\mathbf{e}_i & \mathbf{e}_i & \dots & \mathbf{e}_i
\end{array}\right])
\end{split}
\end{equation}
where $m$ represents the total number of short chunks in the entire audio.
$f$ is
\end{comment}
We employ a stacked RNN with two Bi-LSTM layers followed by two fully-connected layers.
Each Bi-LSTM layer has 512 units (256 forward and 256 backward). The first fully-connected layer has 64 outputs with the ReLU activation, and the second layer has one unit with the sigmoid activation.
% Due to the memory limitation, we calculate the sub-matrices of $\mathbf{S}$ with the size of $64\times64$ and the hop of 32 rather than the entire matrix at both training and inference stage. After obtaining $\mathbf{S}$ by overlap-adding sub-matrices,
After spectral clustering, we obtain the centroids of speakers in the entire long-term audio, which are used as speaker profiles for neural diarization. More details can be found in \citet{LinYLBB19} and \citet{Weiqing2022}.

It should be noted that the number of speaker profiles can be different among long-term audios, which is not friendly to the following neural diarization.
To handle this problem, we zero-pad the profiles of each long-term audio in order to have the same number of profiles $N$.
In our experiments, the maximum number of profiles $N$ is set to 16.
% To perform spectral clustering, we need construct an affinity matrix firstly.
% However, in multi-part meeting scenarios, cosine-based similarity measurement could  there are a lot of overlapping speech regions, which contains more than speaker.

\subsection{Neural Diarization}
Given $T$ acoustic features of a segment $X=[\mathbf{x}_1,\mathbf{x}_2,\dots,\mathbf{x}_T]$ and $N$ speaker profiles of the entire long-term audio $V=[\mathbf{v}_1,\mathbf{v}_2,\dots,\mathbf{v}_N]$, a neural network (NN) is employed to model the posterior probability $y_{n,t}$, which represents whether speaker $n$ talked at timestep $t$:
\begin{equation}
	P(y_{n,t}=1|X,V)=\mathrm{NN}(\mathbf{v}_n, \mathbf{x}_t;X,V)
\end{equation}
We will describe our proposed speaker overlap-aware diarization model in section \ref{sec:osd}.

\subsection{Post-processing}
To obtain the entire transcription with speaker and timestep attributes, a two-step post-processing is performed on the diarization results of clipped segments.
First, we employ a median filter to smooth the segmental results.
According to our preliminary experiments, the filter window size is set to 1.28s.
Then, the smoothed segmental results are concatenated in the chronological order to obtain the time stamps for each speaker.
% We find  which can be used to further extract speaker profiles.
We find that the final performance can be further improved by using the predicted transcriptions to extract profiles. Therefore, we run our system three times iteratively.
% We run embedding extraction, neural diarization and post-processing three times to get the final transcriptions.

\section{Speaker Overlap-aware Neural Diarization}
\label{sec:osd}
\begin{figure}[t!]
	\centering
	\includegraphics[width=0.8\linewidth]{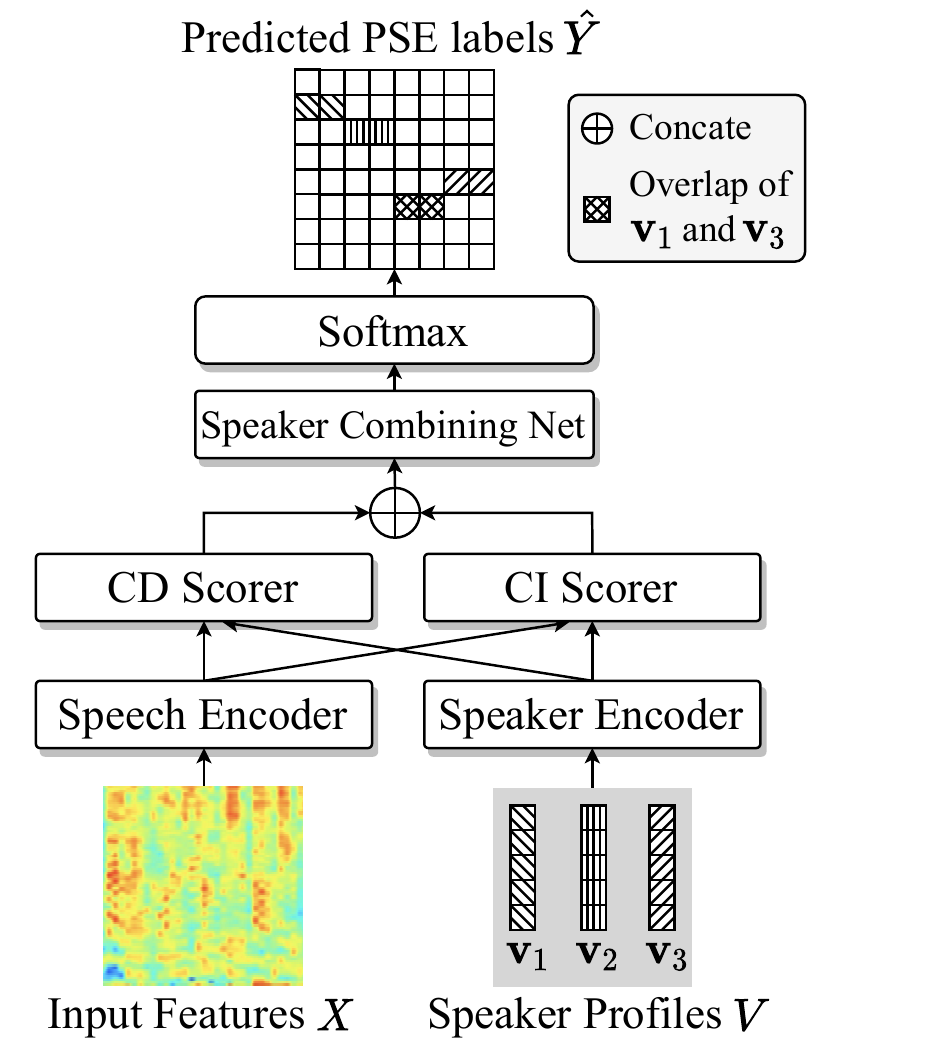}
	\caption{The model diagram of our speaker overlap-aware neural diarization model. CD and CI represent ``context-dependent'' and ``context-independent'', respectively.}
	\label{fig:sond}
\end{figure}

In the proposed speaker overlap-aware neural diarization model (SOND), input acoustic features and speaker profiles are first projected into the same space by the speech and speaker encoders, respectively. Then, the encoded features and profiles are fed into context-dependent and context-independent scorers. Finally, scores of different speakers are combined to predict the power set encoded labels through a speaker combining network. A diagram of SOND is shown in Figure \ref{fig:sond}.

\subsection{Speech and Speaker Encoders}
The speech encoder has the similar model architecture as the embedding extractor, which consists of convolutional blocks (Conv), statistic pooling (SP) and embedding layer (Emb) as described in \ref{sec:app:resnet34}.
However, different from embedding extractor, the statistic pooling of speech encoder is calculated on a window rather then the entire segment:
\begin{equation}
	\begin{split}
		\mathbf{h}_t &= \mathrm{SpeechEncoder}(X) \\
		         &=\mathrm{Emb}(\mathrm{SP}(\mathrm{Conv}(X) \mathbf{I}_{t-2/l:t+l/2}))
	\end{split}	
\end{equation}
where $\mathbf{I}_{t-2/l:t+l/2}$ represents a identity window with ones from $t - l/2$ to $t + l/2$ and zeros otherwise. $\mathbf{h}_t$ denotes the outputs of embedding layer in speech encoder at timestep $t$.
The window length $l$ is set to 20 in our experiments.

The purpose of speaker encoder is to project speaker profiles into the same space of $\mathbf{h}_t$:
\begin{equation}
	\mathbf{\bar{v}}_i=\mathrm{SpeakerEncoder}(\mathbf{v}_i)
\end{equation}
In our experiments, the speaker encoder consists of three fully-connected layers with ReLU activation function, and the output layer has 256 units.
% Residual connections are added between hidden layers to accelerate training process.
\subsection{Context-independent Scorer}
In context-independent (CI) scorer, speakers are detected and tracked through the \emph{global} discriminability, which is learned by contrasting a speaker with others on training set.
We employ the cosine similarity of speech encoding $\mathbf{h}_t$ and projected speaker profile $\mathbf{\bar{v}}_i$ to evaluate the probability of speaker $n$ talking at timestep $t$:
\begin{equation}
	\mathbf{S}^{CI}_{n,t}=\frac{<\mathbf{\bar{v}}_n, \mathbf{h}_t>}{||\mathbf{\bar{v}}_n||_2 ||\mathbf{h}_t||_2}
\end{equation}
where $<\cdot,\cdot>$ represents the inner product of two vectors, and $||\cdot||_2$ denotes the L2 norm.

\subsection{Context-dependent Scorer}
In context-dependent (CD) scorer, speakers are detected and tracked through the \emph{local} discriminability, which is modeled by contrasting a speaker with contextual speakers in the same segment. We employ a multi-head self attention (MHSA) based neural network to predict the context-dependent probabilities of speaker $n$ across all timesteps:
\begin{equation}
\begin{split}
	% &z_{i,0} = \left[\begin{array}{cccc}
	% 	\mathbf{h}_1 & \mathbf{h}_2 & \dots & \mathbf{h}_T \\
	% 	\mathbf{\bar{v}}_i & \mathbf{\bar{v}}_i & \dots & \mathbf{\bar{v}}_i
	% \end{array}\right] \\
	&z_{n,0} = [(\mathbf{h}_1,\mathbf{\bar{v}}_n), (\mathbf{h}_2,\mathbf{\bar{v}}_n),\dots,(\mathbf{h}_T,\mathbf{\bar{v}}_n)] \\
	&\bar{z}_{n,l} = z_{n,l-1} + \mathrm{MHSA}_l(z_{n,l-1}, z_{n,l-1}, z_{n,l-1}) \\
	&z_{n,l} = \bar{z}_{n,l} + \mathrm{max}(0, \bar{z}_{n,l}W^l_1+b^l_1)W^l_2+b^l_2 \\
	&\mathbf{S}^{CD}_{n} = \mathrm{Sigmoid}(W^o z_{n,L^{CD}} + b^{o}) 
\end{split} 
\end{equation}
where $\mathrm{MHSA}_l(Q,K,V)$ represents the multi-head self attention of the $l$-th layer \citep{VaswaniSPUJGKP17} with query Q, key K, and value V matrices.
$W^l_*$ and $b^l_*$ denotes the learnable weight and bias of the $l$-th layer ($o$ for output layer) \footnote{Formally, the weight $W$ of a module should be noted as $W^{module}$, but for notational simplicity, we omit the superscript of different modules.}. 
% For symbol brevity, we omit the subscript 
Detailed model settings of CD scorer are given in Table \ref{tab:cd_scorer}.

\subsection{Speaker Combining Net}
To model speaker dependency and overlaps, we propose the speaker combining network (SCN).
First, we concatenate CI and CD scores across the speaker axis:
\begin{equation}
\begin{split}
	& z^0 = \left[\mathbf{S}_1,\mathbf{S}_2,\dots,\mathbf{S}_T\right] \\
	& \mathbf{S}_t = [\mathbf{S}^{CI}_{t,1}\dots,\mathbf{S}^{CI}_{t,N},\mathbf{S}^{CD}_{t,1},\dots, \mathbf{S}^{CD}_{t,N}]
\end{split}
\end{equation}
where $N$ and $T$ are the maximum numbers of speakers and timesteps, respectively.
Then, the concatenated scores are combined and reassigned through a feed-forward (FF) projection:
\begin{equation}
\begin{split}
	\bar{z}^l &= \mathrm{FF}(z^{l-1}) \\
	    &= \mathrm{LN}(\mathrm{max}(0, z^{l-1}W^l_1+b^l_1))W^l_2 + b^l_2
\end{split}
\end{equation}
where $W^{l=1}_1\in \mathbb{R}^{2N\times d_{ff}}$, $W^{l\ne1}_1\in \mathbb{R}^{N\times d_{ff}}$, $W^l_2\in \mathbb{R}^{d_{ff} \times N}$ and the biases $b^l_1\in \mathbb{R}^{d_{ff}}$, $b^l_2\in \mathbb{R}^{N}$.
$\mathrm{LN}$ represents the layer normalization \citep{BaKH16}.
Subsequently, a sequential memory block \citep{ZhangLYD18} is employed to model the activity sequence of speaker $n$:
\begin{equation}
	z^l_{n,t} = \sum_{i=0}^{L_1}{a^l_{n,i}}\cdot \bar{z}^l_{n,t-i} + \sum_{i=1}^{L_2}{c^l_{n,j}}\cdot \bar{z}^l_{n,t+j} 
\end{equation}
where $a^l_{n}$ denotes the weights of historical items looking back to the past with the size of $L_1$, and ${c^l_{n}}$ represents the weights of look-ahead window into the future with the size of $L_2$.
We stack the FF layer and memory block $L_{SCN}$ times, which is followed by a fully-connected layer to predict the probabilities of $C$ power set encoded labels:
\begin{equation}
	\hat{\mathbf{y}}_t = \mathrm{Softmax}(W^{o}z^{L_{SCN}}_t+b^{o})
\end{equation}
where $W^{o} \in \mathbb{R}^{N \times C}$ and $b^{o} \in \mathbb{R}^{N}$. Softmax activation is performed along the category axis.
Detailed model settings are given in Table \ref{tab:scnet}.

\subsection{Power Set Encoded Labels}
\label{sec:pse}
\begin{figure}[t]
	\centering
	\includegraphics[width=1.0\linewidth]{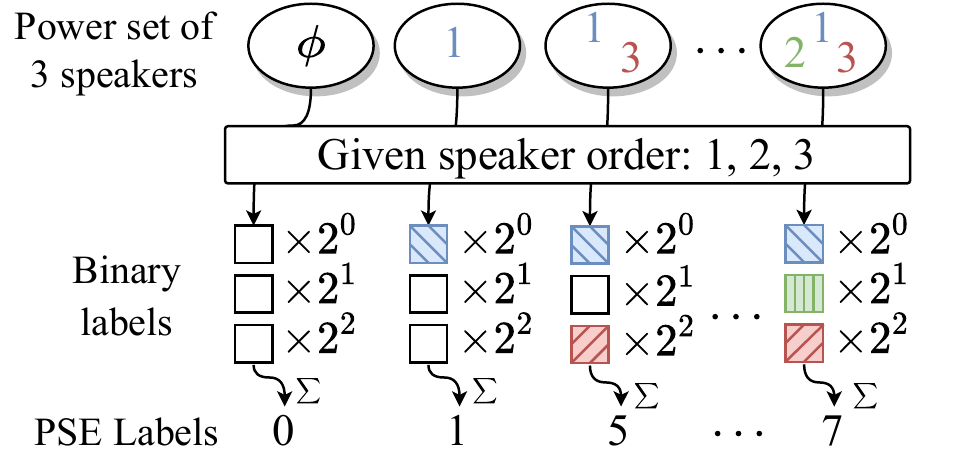}
	\caption{Demonstration of power set encoding with $N=3$ and $K=3$. $\sum$ means the summation operator.}
	% ``0'', ``1'', ``5'' and ``7'' denotes silence, speaker 1, overlap of speaker 1 and 3 and overlap of speaker 1, 2 and 3, respectively
	% $N$ represents the maximum number of speakers in an entire long-term audio. $K$ denotes the maximum number of overlapped speakers at a single timestep.
	\label{fig:pse}
\end{figure}

% In previous studies, overlapping speech diarization is always formulated as a multi-label classification problem by treating each speaker as a category, where the binary label $y_{n,t}$ indicates whether speaker $n$ talks at timestep $t$.
% In this formulation, the correlation between speakers is ignored, and a threshold is needed to obtain the final diarization results. To overcome the above issues, 
In this section, we reformulate overlapping speech diarization as a single-label prediction problem through a power set, which can model speaker overlaps explicitly.
Given $N$ speakers $\{1,2,\dots,N\}$, their power set (PS) is defined as follows:
\begin{equation}
	\begin{split}
		\mathrm{PS}(N) &=\{A|A\subseteq \{1,2,\dots,N\}\} \\
		&=\{\phi,\{1\},\{2\},\{1,2,n,\}, \dots \}         
	\end{split}
\end{equation}
where $\phi$ means the empty set. 
From the definition, we can see that each element of PS represents a combination of speakers, and the power set contains all possible combinations. 
Therefore, if we treat the elements of PS as classification categories, an overlapping speech frame can be uniquely assigned with a single label.
% We employ a simple approach to encode the PS elements, where 
Power-set encoded (PSE) label $\tilde{y}_t$ is obtained by treating the binary label $y_{n,t}$ as an indicator variable:
\begin{equation}
	\tilde{y}_t= \sum_{n=1}^{N}{y_{n,t}\cdot 2^{n-1}}
\end{equation}
% where $\delta(n, S)$ is the Dirac function, which equals one if speaker $n$ belongs to $S$ and zero otherwise.
% For example, the speech encoding $h_t$ contains speaker \emph{1} and \emph{3}, and the speaker order in $V$ is [\emph{1}, \emph{2}, \emph{3}]. Then, the power set encoded label $y_t$ will be $5(1\times2^0+0\times2^1+1\times2^2)$.
By applying power-set encoding on $N$ speakers, we get $2^N$ categories, which may be impractical for a large number of speakers. 
Fortunately, the maximum number of overlapping speakers $K$ is always small (e.g., two, three or four at most) in real-world applications. Therefore, the number of reasonable categories can be reduced to:
\begin{equation}
	C=\sum_{k=0}^{K}{N \choose k} = \sum_{k=0}^{K}\frac{N!}{k!(N-k)!}
	\label{eq:pse}
\end{equation}
In this way, overlapping speech diarization is reformulated as a single-label prediction problem with $C$ categories.
% In our experiments, we set $N=16$ and evaluate the impact of $K$ on diarization performance.
Figure \ref{fig:pse} shows an example of PSE.

\subsection{Training Objective}
\label{sec:loss}
We adapt a multi-task learning strategy to optimize our SOND model.
The main training objective is minimizing the cross entropy loss between predicted probabilities of PSE labels $\hat{\mathbf{y}}_t$ and their ground-truth counterparts $\tilde{y}_t$:
\begin{equation}
	\mathcal{J}^{CE}(\theta) = \frac{1}{T}\sum_{t=1}^{T}{\mathrm{CrossEntropy}(\hat{\mathbf{y}}_t, \tilde{y}_t)}
\end{equation}
where $\theta$ denotes learnable parameters of SOND.
The second training objective is minimizing the similarity between projected speaker profiles $\bar{\mathbf{v}}_n$ in an entire long-term audio:
\begin{equation}
	\mathcal{J}^{sim}(\theta) = \sum_{i,j=1}^{N}{\mathrm{max}\left(0,\frac{<\bar{\mathbf{v}}_i, \bar{\mathbf{v}}_j>}{||\bar{\mathbf{v}}_i||_2 ||\bar{\mathbf{v}}_j||_2}+\delta-1\right)}
	\label{eq:simi-loss}
\end{equation}
where $\delta$ represents the expected margin of different speaker profiles.
The total training objective is obtained as follows:
\begin{equation}
	\theta = \arg\min_{\theta} \mathcal{J}^{CE}(\theta) + \lambda \mathcal{J}^{sim}(\theta)
\end{equation}
where $\lambda$ is a hype-parameter to balance the CE and similarity loss. According to our preliminary experiments, we find $\lambda$ slightly affects the final results, thus we simply set it to 1 in this paper.
\section{Experiments}
\label{sec:exp}
\subsection{Alimeeting Dataset}
We conduct experiments on the AliMeeting corpus \citep{YuZFXZDHGYMXB22}, which includes far-field long-term audios recorded by 8-channel microphone array in real meeting scenarios. 
Since we focus on monaural speaker diarization in this paper, the model training and evaluation are all based on the first-channel data.
The training (Train) set of AliMeeting contains 212 audios, about 104.75 hours.
The evaluation (Eval) set contains 8 audios (about 4 hours), which are used for model selection and hyper-parameter tuning.
The test set (Test) contains 20 audios (about 10 hours), which are employed to evaluate model performance.
Each audio consists of a 15 to 30-minute meeting with 2 to 4 speakers.
Note that speakers in Train, Eval and Test sets are different from each other.
We provide more statistics of AliMeeting in \ref{sec:app:alimeeting}.
% To highlight the speaker overlap, the sessions with 4 participants account for 59\%, 50\% and 57\% sessions in Train, Eval and Test, respectively.

\subsection{Data for Training Embedding Extractor}
We first pre-train the speaker embedding extractor with utterances from the CN-Celeb corpus \citep{FanKLLCCZZCW20}.
CN-Celeb is a large-scale speaker recognition dataset collected ``in the wild''. 
This dataset contains 274 hours from 1,000 celebrities. 
% It covers 11 genres and the total duration of utterances is about 274 hours.
To enrich training samples in CN-Celeb, we perform data augmentation with the MUSAN noise dataset \citep{musan2015} and simulated room impulse responses (RIRs) \citep{KoPPSK17}. We first perform the amplification and tempo perturbation (change audio playback speed but do not change its pitch) on speech. Then, 40,000 simulated RIRs from small and medium rooms are used for reverberation. Finally, the reverberated signals are mixed with background noises at the speech-to-noise rates (SNRs) of 0, 5, 10, and 15 dB.

After pre-training with CN-Celeb, we finetune the embedding extractor on AliMeeting.
% To train a speaker embedding extractor, a training utterance must contain a single speaker in it.
As the AliMeeting corpus does not provide single-speaker utterances, we select non-overlapping segments according to ground-truth transcriptions, where segments shorter than two seconds are dropped.
As a result, we obtain 118,350 utterances from Train set for training and 1,833 utterances from Eval set for parameter selection.
The model with lowest equal error rate on Eval set is employed to extract speaker embeddings.
%  and initialize the speech encoder of SOND.

\subsection{Data for Training SOND}
We first pre-train SOND with a simulated dataset created from the AliMeeting Train set. Details of simulation process are provided in \ref{sec:app:simulation}.
We run the simulation process 450,000 times to obtain enough training samples.
After pre-training with simulated data, we further finetune SOND with real segments from AliMeeting Train set.
% We first remove silence regions in long-term audios according to VAD.
% Then, we uniformly clip the audios and labels into segments with the duration of 16s and the shift of 4s.
Segments from the same long-term audio have the same speaker profiles, which contain all speakers talking in the audio. 
As a result, we obtain 92,569 real training samples.
The Eval and Test sets are processed in the same manners as Train set.

\subsection{Baselines}
We compare our method with VBx and TSVAD \cite{MedennikovKPKKS20}.
VBx is a clustering-based algorithm, which achieves promising results on several diarization tasks \citep{landini2022bayesian}.
We reuse the codes released along with AliMeeting to implement VBx\footnote{Code is available at https://github.com/yufan-aslp\\/AliMeeting}.
% Similar to SOND, TSVAD is also a neural diarization model.
We adopt the same model architecture as described in \citet{Weiqing2022} to implement TSVAD.
% , and use it to replace SOND as our second baseline.
Note that TSVAD is a strong baseline, which achieves the state-of-the-art performance on the AliMeeting corpus \citep{YuZGFDZHXTWQLYM22}.

\subsection{Experimental Settings}
We implement and train our models with TensorFlow 1.12.
Input features are generated by 80-dimensional Mel-frequency filter-banks on each frame, with a window size of 25ms shifted every 10ms.
Models are trained with Adam optimizer.
% The maximum speaker number $N$ is set to 16, and 
The maximum number of overlapped speakers $K$ is set to 4.
As a result, PSE labels have 2,517 possible categories.

The embedding extractor is first pre-trained for 200,000 steps on CN-Celeb with the learning rate of 1e-4 and the batch size of 64.
Then, we finetune it for another 100,000 steps on AliMeeting with the learning rate of 1e-5.

The training procedure of SOND has three stages.
At the first stage, the speech encoder is frozen and initialized with the pre-trained parameters of embedding extractor.
We use the simulated data to train the remaining parameters for 200,000 steps with an initial learning rate of 1.0 and 10,000 warm-up steps.
At the second stage, we unfreeze the speech encoder and train the whole SOND model on simulated data for 400,000 steps with a fixed learning rate of 1e-4.
At the third stage, we employ real training data to finetune the whole model for 50,000 steps with a fixed learning rate of 1e-5.
The batch size is set to 32 for all stages.
Five models with the best performance on Eavl set are averaged and evaluated on the Test set.

Performance are measured by diarization error rate (DER) \citep{FiscusAMG06}, which is a commonly-used metric for speaker diarization tasks \footnote{The toolkit of DER calculation is available at https://\\github.com/nryant/dscore.}. See \ref{sec:app:der} for more details of DER.

\begin{comment}
\multicolumn{3}{l}{\textbf{Multi-channel \& Multi-model}} \\
\hline & & \\[-2.2ex]
\citet{ZhengLWMKWWSM22} & \textbf{3.22} & \textbf{3.98} \\
\citet{ShenLFWWTZYM22}& 5.79 & 7.23 \\
\hline & & \\[-2.2ex]
\multicolumn{3}{l}{\textbf{Multi-channel \& Single-model}} \\
\hline & & \\[-2.2ex]
\citet{Weiqing2022} & \textbf{2.26} & \textbf{2.98} \\
\citet{HeLZYZWNCLDL22} & 2.82 & 4.05 \\
\citet{ZhengLWMKWWSM22} & 3.64 & 5.63 \\
\hline & & \\[-2.2ex]
\multicolumn{3}{l}{\textbf{Single-channel \& Single-model}} \\
\hline & & \\[-2.2ex]
\end{comment}

\section{Results}
\begin{table}[t!]
	\centering
	\setlength{\tabcolsep}{1.9mm}
	\begin{tabular}{lcc}
		\toprule
		\textbf{System} & \textbf{Eval} & \textbf{Test} \\
		\hline & & \\[-2.2ex]
		VBx \citep{DiezBWRC19} & 15.24 & 15.60 \\
		% Spectral Clustering & 14.49 & 14.71 \\
		IV-TSVAD \citep{ZhengLWMKWWSM22} & 5.46 & 6.92 \\
		SC-TSVAD \citep{Weiqing2022} & 3.49 & - \\
		PSE+TSVAD & 3.12 & 4.76 \\
		PSE+SOND (Ours) & \textbf{2.70} & \textbf{4.46} \\
		\bottomrule 
	\end{tabular}
	\caption{The diarization performance of different methods in terms of DER (\%) on the AliMeeting Eval and Test sets. Best results are highlighted by boldface.}
	\label{tab:overall_der}
\end{table}
% For fair comparison, we divide current methods into three groups.
% In a ``Multi-channel \& Multi-model (MCMM)'' method, the multi-channel data is fed into several models, and their outputs are fused to get the final results.
% In a ``Multi-channel \& Single-model (MCSM)'' method, there is only a single model to process the multi-channel data.
% In a ``Single-channel \& Single-model (SCSM)'' method, the single-channel data is fed into a single model to obtain the diarization results.
% Our system belongs to the SCSM group.
Table \ref{tab:overall_der} shows the diarization performance of different methods in terms of DER on Eval and Test sets.
As expected, VBx achieves the worst performance, which is because the overlap ratio of AliMeeting Test set is very high (42.8\%), and VBx can not deal with speaker overlaps.
By involving a TSVAD model to handle overlap, IV-TSVAD and SC-TSVAD achieve much better performance than VBx.
Our implemented TSVAD baseline outperforms the two methods, although they have similar model architectures. 
This indicates the effectiveness of proposed formulation and PSE.
By replacing TSVAD with the proposed SOND, our system achieves the best performance with 13.46\% and 6.30\% relative improvements on Eval and Test set.
% As a results, our method achieves the best performance compared with other SCSM methods.

% Compared with MCMM and MCSM methods, our system outperforms \citet{ShenLFWWTZYM22} and the single-system version of \citet{ZhengLWMKWWSM22}. %, which utilize the multi-channel data.
% Compared with \citet{HeLZYZWNCLDL22}, the performance of our method is better on Eval set but worse on Test set. 
% This is because the overlap ratio of Test set is higher, and the multi-channel information becomes more important under high-overlap situations.
% The performance gap between our system and \citet{Weiqing2022} inspires us to extend SOND with spatial information from multi-channel data.
% comparision with baselines
\subsection{Hyper-Parameter Tuning}
\begin{table}[t!]
	\centering
	\setlength{\tabcolsep}{3.0mm}
	\begin{tabular}{lcccc}
		\toprule
		\textbf{Model ID} & \textbf{\emph{K}} & \textbf{\emph{C}} & \textbf{Eval} & \textbf{Test} \\
		\hline & & & & \\[-2.2ex]
		SOND-K1 & 1 & 17 & 15.69 & 16.34 \\
		SOND-K2 & 2 & 137 & 4.06 & 6.15 \\
		SOND-K3 & 3 & 697 & 2.97 & 4.63 \\
		SOND-K4 & 4 & 2517 & 2.70 & 4.46 \\
		\bottomrule
	\end{tabular}
	\caption{Hyper-parameter tuning on the maximum number of overlapped speakers $K$. The performance is measured by DER(\%) on Eval and Test set. $C$ denotes the number of PSE categories.}
	\label{tab:k_tune}
\end{table}
We perform hyper-parameter tuning on the maximum number of overlapped speakers $K$, and the results are shown in Table \ref{tab:k_tune}.
Since the audios of AliMeeting corpus consist of 4 speakers at the most, we evaluate 1, 2, 3 and 4 in this experiment.
As expected, SOND-K1 achieves the worst performance on both Eval and Test set, which is because speaker overlap is ignored in SOND-K1 as VBx does.
By setting $K$ to 2, speaker overlap are considered and diarization errors are significantly reduced.
When $K$ increases, more speaker combinations are encoded, and diarization performance is further improved.
% Although the number of PSE categories increases exponentially while $K$ increasing, $K=4$ is a large enough setting for most real meeting scenarios.
According to the results, we set $K=4$ in all following experiments.
% Even a human can not identify speakers from a mixture generated by more than 4 speakers.

\begin{table}[t!]
	\centering
	\setlength{\tabcolsep}{4.0mm}
	\begin{tabular}{lccc}
		\toprule
		\textbf{Model ID} & \textbf{Margin} $\delta$ & \textbf{Eval} & \textbf{Test} \\
		\hline & & & \\[-2.2ex]
		SOND-S & 0.0 & 2.79 & 5.64 \\
		SOND-M & 0.5 & 2.73 & 4.79 \\
		SOND-L & 1.0 & 2.70 & 4.46 \\
		\bottomrule
	\end{tabular}
	\caption{Hyper-parameter tuning on the similarity margin $\delta$. The performance is measured by DER(\%) on Eval and Test set.}
	\label{tab:delta_tune}
\end{table}
% for similarity margin $\delta$ in equation (\ref{eq:simi-loss}).
We also evaluate the impact of similarity margin $\delta$ in Eq.\ref{eq:simi-loss}.
Results of $\delta=0.0$, $0.5$ and $1.0$ are shown in Table \ref{tab:delta_tune}.
Margins larger than 1.0 are excluded, since they indicate two speaker embeddings are negatively correlated, which is not desired here.
With the smallest margin, SOND-S achieves the worst performance in terms of DER. 
By increasing the similarity margin from 0.0 to 0.5, SOND-M achieves a 15.07\% relative improvement on Test set.
Further increasing the margin to 1.0 brings another 6.89\% relative error reduction on Test set (seen in SEND-L).
We find that increasing the margin does not much improve the performance on Eval set.
The reason is two-fold. First, models are selected with Eval set, which may be over-tuned.
Second, the overlap ratio of Test set is higher than that of Eval set, which places more demands on global speaker discriminability.

\section{Analysis}
\subsection{Ablation Study}
\begin{table}[t!]
	\centering
	\setlength{\tabcolsep}{3.0mm}
	\begin{tabular}{lcccc}
		\toprule
		\textbf{Model} & \textbf{MD} & \textbf{FA} & \textbf{SC} & \textbf{DER} \\
		\hline & & & &  \\[-2.2ex]
		SOND & 2.29 & 1.17 & 1.02 & 4.46 \\
		$-$ Spk. Enc. & 2.64 & 1.55 & 1.26 & 5.43 \\
		$-$ CD scorer & 2.49 & 1.90 & 1.88 & 6.25 \\
		$-$ CI scorer & 2.39 & 1.13 & 1.09 & 4.60 \\
		$-$ PSE & 2.29 & 1.78 & 0.95 & 5.00 \\
		$-$ SCN & 2.56 & 1.90 & 1.42 & 5.88 \\
		\bottomrule
	\end{tabular}
	\caption{Ablation study on components of SOND. Results are measured in percentage (\%). MD, FA and SC denote errors of miss detection, false alarm and speaker confusion, respectively. Spk. Enc. represents speaker encoder for short.}
	\label{tab:ablation}
\end{table}
We conduct the ablation and replacement study to evaluate each component of SOND. 
Results are shown in Table \ref{tab:ablation}.
We find removing speaker encoder leads to a 21.75\% relative degradation on DER, which reveals the importance of speaker encoder in overlapped regions.
While removing CD scorer from SOND leads to a significant performance degradation, removing CI scorer only causes a slight impact.
%  is observed, which indicates the local discriminability is very important for speaker diarization.
% On the contrary,  Despite this, CI scorer still brings a 3.04\% relative improvement.
Replacing PSE labels with binary multi-labels makes the overall DER increase from 4.46\% to 5.00\%, which reveals the effectiveness of PSE.
It is interesting to find that PSE affects FA errors severely. This is because PSE explicitly models the dependency of speakers by encoding their combinations.
In this way, unrelated speakers will be excluded for an activated speaker, resulting in much less false alarm errors.
The impact of SCN is shown in the last row of Table \ref{tab:ablation}.
% Removing SCN from SOND leads to a significant performance degradation. 
This result reveals the necessity of modeling speaker dependency through combining and reassignment. 
% We further analysis the effectiveness of SCN in the next section.

\begin{table}[t!]
	\centering
	\setlength{\tabcolsep}{1.5mm}
	\begin{tabular}{lcccc}
		\toprule
		\textbf{Model} & \textbf{Layers} & \textbf{Units} & \textbf{Size(M)} & \textbf{DER(\%)} \\
		\hline & & & &  \\[-2.2ex]
		SCN & 6 & 512 & 19.08 & 4.46 \\
		\hline & & & &  \\[-2.2ex]
		None & - & - & 14.35 & 5.88 \\
		FCN & 6 & 1024 & 20.45 & 5.69 \\
		CNN & 6 & 512 & 19.29 &  5.04 \\
		BiLSTMP & 4 & 512 & 19.30 & 4.73 \\
		\bottomrule
	\end{tabular}
	\caption{Diarization performance of different models for speaker combining on Test set.}
	\label{tab:scn}
\end{table}
\subsection{SCN vs. Other Model Architectures}
\label{sec:scn_vs}
To evaluate the effectiveness of SCN, we compare it with other commonly-used network architectures, such as fully-connected networks (FCN), convolutional neural networks (CNN) and bi-directional LSTM with projection (BiLSTMP).
For fair comparison, we keep other components of SOND unchanged and tune model settings to make them have the same size. 
Table \ref{tab:scn} shows the model settings and comparison results.
We find removing SCN from SOND causes a significant performance degradation.
By employing a FCN to combine frame-level speaker activities, only a slight improvement is obtained. This indicates that frame-level information is not sufficient for speaker combining.
Compared with FCN, CNN achieves a much lower DER, which reveals the importance of sequential modeling.
To further enhance sequential modeling, BiLSTMP outperforms CNN with a 6.15\% relative improvement.
With the similar model size, the proposed SCN achieves better performance than BiLSTMP in terms of DER.
In addition, outputs of SCN can be calculated in parallel, which is more friendly to modern computers than BiLSTMP.

\subsection{Sensitivity to Initial Speaker Profiles}
Sensitivity to initial speaker profiles is an important property for neural diarization models in real-world applications.
We compare the sensitivity of our SOND and the baseline TSVAD in Table \ref{tab:profile_type}, where ``Clustering'' means profiles are obtained with the results of spectral clustering (SC) and ``Oracle'' means profiles are extracted using ground-truth transcriptions. For convenience, we also provide the diarization performance of SC in the table, which reflects the quality of speaker profiles.

As expected, models using oracle profiles achieve better performance than those with clustering-based profiles.
The performance gap is more significant on Test set, which is mainly due to the over-tuning issue on Eval set and the higher overlap ratio of Test set. 
Therefore, we mainly focus on the results of Test set.
By replacing oracle profiles with clustering-based ones, an 8.67\% relative degradation of TSVAD is observed.
For our SOND, using clustering-based profiles leads to a 5.94\% relative degradation, which is smaller than TSVAD.
This indicates that the proposed SOND is less sensitive to speaker profiles than TSVAD. 

To dig out the reason of SOND's robustness to noises in profiles, we further perform an ablation study.
From the results, we can see that the speaker encoder and CI scorer are most important components.
Removing them causes 23.97\% and 10.05\% relative degradation.
This reveals that global speaker discriminability is crucial for the robustness to noises in profiles.
\begin{table}[t!]
	\centering
	\setlength{\tabcolsep}{1.0mm}
	\begin{tabular}{lcccc}
		\toprule
		\textbf{Model} & \textbf{Profile Type} & \textbf{Eval} & \textbf{Test} & \textbf{Deg.} \\
		\hline & & & & \\[-2.2ex]
		SC & - & 14.49 & 14.71 & -\\
		\hline & & & & \\[-2.2ex]
		TSVAD & Oracle & 3.07 & 4.38 & - \\
		TSVAD & Clustering & 3.12 & 4.76 & 8.68 \\
		SOND & Oracle & 2.68 & 4.21 & - \\
		SOND & Clustering & 2.70 & 4.46 & 5.94 \\
		\hline & & & & \\[-2.2ex]
		$-$ CI scorer & Oracle & 3.02 & 4.18 & - \\
		$-$ CI scorer & Clustering & 3.03 & 4.60 & 10.05 \\
		$-$ CD scorer & Oracle & 3.31 & 5.81 & - \\
		$-$ CD scorer & Clustering & 3.41 & 6.25 & 7.57 \\
		$-$ Spk. Enc. & Oracle & 2.97 & 4.38 & - \\
		$-$ Spk. Enc. & Clustering & 2.97 & 5.43 & 23.97 \\
		\bottomrule
	\end{tabular}
	\caption{The sensitivity of models to profile types. Performance is measured by DER(\%) on Eval and Test sets. ``Deg.'' represents relative performance degradation on Test set by replacing oracle profiles with clustering-based ones. }
	\label{tab:profile_type}
\end{table}
\section{Conclusion and Future Work}
In this paper, a hybrid diarization system improved by PSE and SOND is proposed for long-term audios in multi-party meeting scenarios.
Specifically, we perform spectral clustering on the neural network based affinity matrix to extract speaker profiles.
Then, the profiles are consumed by SOND to predict PSE labels at different time-steps.
We find that explicitly modeling speaker dependency and overlaps via PSE labels much reduces diarization errors.
In SOND, the local speaker discriminability, discovered by CD scorer, is important for final diarization performance. Meanwhile, the global speaker discriminability, modeled by CI scorer, can much improve the robustness to noises in speaker profiles.
Compared with other network architectures, SCN is more efficient for combining and reassigning speaker activities.
As a result, our system outperforms the state-of-the-art monaural speaker diarization methods with a 6.30\% relative improvement.
% there is still a performance gap to multi-channel models.
In the future, we are interested on extending our SOND model to leverage spatial information for multi-channel data.
% we employ the power set encoded labels to reformulate overlapped speaker diarization from a multi-label classification problem into a single-label prediction problem, in which speaker overlaps and dependency can be explicitly modeled. To fully leverage this formulation, we propose the SOND model, which consists of a CI scorer, CD scorer, 
% where a CI scorer is used to model global speaker discriminability and a CD scorer is employed to discover local speaker discriminability. The 

\section*{Limitations}
Similar to other neural diarization methods, the proposed method has two main limitations.
First, the model architecture of SOND is dependent on the maximum number of speakers in a long-term audio.
To deal with variable speaker numbers of different audios, a fixed number $N$ is set and profiles are zero-padded. 
This can cause extra computational cost, especially when the real speaker number is much smaller than $N$.
Second, power set encoding has a limited scalability to very large number of speakers.
According to equation (\ref{eq:pse}), we can see that the number of PSE categories increases exponentially with the maximum number of speakers.
When $N$ is small ($\leqslant 16$), setting a maximum number of overlapped speakers can alleviate this problem. 
For the massively ($N>16$) multi-party meeting scenarios, this limitation is still an obstacle.

\bibliography{anthology,custom}
\bibliographystyle{acl_natbib}

\appendix
\section{Appendix}
\subsection{Speaker Embedding Extractor}
\label{sec:app:resnet34}
The architecture details of our speaker embedding extractor is given in Table \ref{tab:resnet34}, where $T$ and $D$ denote the sequence length and dimension of input features, respectively. $C$ represents the total number of speakers on the training set. ``ds'' means down-sampling layer for short.
\begin{table}[htb]
	\centering
	\setlength{\tabcolsep}{0.6mm}
	\begin{tabular}{l|c|c}
		\toprule
		\textbf{Layer} & \textbf{Parameters} & \textbf{Output} \\
		\hline & & \\[-2.5ex]
		reshape & - & $T \times D \times 1$ \\[0.1ex]
		\hline
		conv1 & 3 $\times$ 3, 32, 1 & $T \times D\times$ 32 \\
		\hline & & \\[-2.4ex]
		conv2\_x &  $\left[ \begin{array}{c}
			3 \times 3, 32, 1 \\
			3 \times 3, 32, 1
		\end{array}\right] \times 3$ & $T \times D \times$ 32 \\[2.0ex]
		\hline & & \\[-2.4ex]
		ds1 & 3 $\times$ 3, 64, 2 & $\frac{T}{2}\times \frac{D}{2} \times 64$ \\[0.4ex]
		\hline & & \\[-2.4ex]
		conv3\_x &  $\left[ \begin{array}{c}
			3 \times 3, 64, 1 \\
			3 \times 3, 64, 1
		\end{array}\right] \times 4$ & $\frac{T}{2}\times \frac{D}{2} \times 64$  \\[2.0ex]
		\hline & & \\[-2.4ex]
		ds2 & 3 $\times$ 3, 128, 2 & $\frac{T}{4}\times \frac{D}{4} \times 128$ \\[0.4ex]
		\hline & & \\[-2.4ex]
		conv4\_x &  $\left[ \begin{array}{c}
			3 \times 3, 128, 1 \\
			3 \times 3, 128, 1
		\end{array}\right] \times 6$ & $\frac{T}{4}\times \frac{D}{4} \times 128$  \\[2.0ex]
		\hline & & \\[-2.4ex]
		ds3 & 3 $\times$ 3, 256, 2 & $\frac{T}{8}\times \frac{D}{8} \times 256$ \\[0.4ex]
		\hline & & \\[-2.4ex]
		conv5\_x &  $\left[ \begin{array}{c}
			3 \times 3, 256, 1 \\
			3 \times 3, 256, 1
		\end{array}\right] \times 3$ & $\frac{T}{8}\times \frac{D}{8} \times 256$  \\[2.0ex]
		\hline & & \\[-2.4ex]
		fc & $256 \times 256$, - , - & $\frac{T}{8}\times \frac{D}{8} \times 256$  \\[0.4ex]
		\hline & & \\[-2.4ex]
		pooling & global statistic pooling & $1 \times 512$ \\[0.4ex]
		\hline & & \\[-2.4ex]
		embedding & $512 \times 256$, -, - & $1 \times 256$ \\[0.4ex]
		\hline & & \\[-2.4ex]
		output & $256 \times M$, softmax & $1 \times C$ \\
		\bottomrule 
	\end{tabular}
	\caption{The architecture of our speaker embedding extractor. Parameter settings are given in the format of ``height $\times$ width, channel number, stride''.}
	\label{tab:resnet34}
\end{table}

\subsection{Detailed Model settings}
\label{sec:app:scn}
The model settings of CD scorer and SCN are given in Table \ref{tab:cd_scorer} and Table \ref{tab:scnet}.
\begin{table}[thb]
	\centering
	\setlength{\tabcolsep}{8.0mm}
	\begin{tabular}{lr}
		\hline
		\textbf{Name} & \textbf{Value} \\
		\hline & \\[-2.2ex]
		Layers $L^{CD}$ & 4 \\
		Attention Dimension & 512 \\
		Attention Heands & 4 \\
		Weight Dimension & 1024 \\
		Output Dimension & 1 \\
		\hline 
	\end{tabular}
	\caption{Detailed model settings of CD scorer.}
	\label{tab:cd_scorer}
\end{table}

\begin{table}[thb]
	\centering
	\setlength{\tabcolsep}{6.0mm}
	\begin{tabular}{lr}
		\hline
		\textbf{Name} & \textbf{Value} \\
		\hline & \\[-2.2ex]
		Layers $L^{SCN}$ & 6 \\
		FF Dimension $d_{ff}$ & 512 \\
		Look-back Length $L_1$ & 15 \\
		Look-ahead Length $L_2$ & 15 \\
		\hline 
	\end{tabular}
	\caption{Detailed model settings of SCN.}
	\label{tab:scnet}
\end{table}

\subsection{Details of AliMeeting dataset}
\label{sec:app:alimeeting}
We use open-source AliMeeting corpus for our experiments, which is available at https://www.openslr.org/119. Table \ref{tab:alimeeting} shows the
statistics of the dataset.
\begin{table}[htb]
	\centering
	\setlength{\tabcolsep}{2.0mm}
	\begin{tabular}{lccc}
		\toprule
		 \textbf{Attributes}& \textbf{Train} & \textbf{Eval} & \textbf{Test} \\
		\hline & & & \\[-2.2ex]
		Duration (hour)  & 104.75 & 4.00 & 10.00 \\
		Sessions & 212 & 8 & 20 \\
		Rooms & 12 & 5 & 6 \\
		Total Speakers & 456 & 25 & 60 \\
		Total Males & 246 & 12 & 31 \\
		Total Females & 210 & 13 & 29 \\
		Overlap Ratio (\%) & 42.27 & 34.20 & 42.8 \\
		\bottomrule 
	\end{tabular}
	\caption{Statistics of AliMeeting corpus.}
	\label{tab:alimeeting}
\end{table}
\subsection{Data Simulation Process}
\label{sec:app:simulation}
The simulated training samples for SOND are created as follows:
\begin{enumerate}
	\item Select all non-overlapped speech for each speaker in the Train set.
	\item Extract the binary labels from the transcriptions and remove the silence regions.
	\item Randomly select a continuous segment of binary labels with the duration of 16s and fill the active region with non-overlapped speech segments.
	\item Extract speaker profiles for used speakers in this segment.
	\item Augment speaker profiles with unused speakers in the training set.
	\item Repeat Step 3-5 many times.
\end{enumerate}

\subsection{Evaluation Metric}
\label{sec:app:der}
Diarization error rate (DER) is calculated as: the summed time of three different errors of speaker confusion (SC), false alarm (FA) and missed detection (MD) divided by the total duration time:
\begin{equation}
	\mathrm{DER}=\frac{\mathcal{T}_{SC}+\mathcal{T}_{FA}+\mathcal{T}_{MD}}{\mathcal{T}_{Total}} \times 100\%
\end{equation}
where $\mathcal{T}_{SC}$, $\mathcal{T}_{FA}$ and $\mathcal{T}_{MD}$ are the duration of the three errors, and $\mathcal{T}_{Total}$ is the total duration.
In order to mitigate the effect of inconsistent annotations and human errors in reference transcriptions, we set a 0.25 second ``no score'' collar around every boundary of the reference segment.

\subsection{Average Runtime}
\begin{table}[htb]
	\centering
	\setlength{\tabcolsep}{4.0mm}
	\begin{tabular}{lrc}
		\toprule
		\textbf{Stage} & \textbf{Training} & \textbf{Inference} \\
		\hline & & \\[-2.2ex]
		 Pre-train & 1d 17.8h & \multirow{3}{*}{3min} \\
		 Train & 3d 22.9h & \\
		 Finetune & 11.4h & \\
		\bottomrule 
	\end{tabular}
	\caption{Average runtime of SOND.}
	\label{tab:runtime}
\end{table}
In Table \ref{tab:runtime}, we list the average runtime using two V100 GPUs of 1) Pre-train: freezing speech encoder and training the remaining parameters on simulated data, 2) Train: training the whole model on simulated data, 3) Finetune: finetuning the whole model on real data. The cost time of inference on Test set is also given in the table.

\subsection{Computing Infrastructure}
We conduct our experiments on NVIDIA V100 GPU (16GB) and Intel(R) Xeon(R) Platinum 8163 32-core CPU @ 2.50GHz.

\end{document}